\documentclass[acmsmall]{acmart}
\AtBeginDocument{%
  \providecommand\BibTeX{{%
    \normalfont B\kern-0.5em{\scshape i\kern-0.25em b}\kern-0.8em\TeX}}}

\setcopyright{rightsretained}
\acmJournal{PACMHCI}
\acmYear{2025} \acmVolume{9} \acmNumber{2} \acmArticle{CSCW174} \acmMonth{4} \acmPrice{}\acmDOI{10.1145/3711072}

\usepackage{enumitem}
\acmSubmissionID{1357}

\begin{document}

\newcommand{\systemname}{\textsc{Rx RiskMap}}

%%
%% The "title" command has an optional parameter,
%% allowing the author to define a "short title" to be used in page headers.
% \title{Designing ML-Driven Overdose Risk Visualizations for Local Public Health Decision-Making}

% \title{Communicating Risk of Fatal Overdoses using Machine Learning and Visualization for Local Public Health Decision-Making}

% \title{Designing a Dashboard for the Opioid Epidemic:  How Does Machine Learning and Visualization Fit into Local Public Health Decision-Making?}

% \title{Designing ML-Driven Overdose Risk Visualizations for Local Public Health Decision-Making}

% \title{The Role of Predictions and Visualizations in Local Public Health Decision-Making for the Opioid Epidemic}

% \title{Predicting and Visualizing Overdose Risk to Support Local Public Health Decision-Making}
% \title[Static Algorithms in an Evolving Epidemic]{Static Algorithms in an Evolving Epidemic: \edit{Understanding the Potential of Human-AI Risk Prediction to Support Regional Overdose Prevention}}
\title[Static Algorithm, Evolving Epidemic]{Static Algorithm, Evolving Epidemic: Understanding the Potential of Human-AI Risk Assessment to Support Regional Overdose Prevention}

% Centralized Data, Local Decisions
% Aggregate Algorithms, Local Decisions
% Static Algorithms, Evolving Decisions

%%
%% The "author" command and its associated commands are used to define
%% the authors and their affiliations.
%% Of note is the shared affiliation of the first two authors, and the
%% "authornote" and "authornotemark" commands
%% used to denote shared contribution to the research.
\author{Venkatesh Sivaraman}
\email{venkats@cmu.edu}
\orcid{0000-0002-6965-3961}

\affiliation{%
  \institution{Carnegie Mellon University}
  \city{Pittsburgh}
  \state{Pennsylvania}
  \country{USA}
}

\author{Yejun Kwak}
\orcid{0009-0006-4582-8938}
\affiliation{%
  \institution{Carnegie Mellon University}
  \city{Pittsburgh}
  \state{Pennsylvania}
  \country{USA}
}

\author{Courtney Kuza}
\orcid{0000-0001-9137-0928}
\affiliation{%
% \department{Center for Pharmaceutical Policy and Prescribing}
  \institution{University of Pittsburgh}
  \city{Pittsburgh}
  \state{Pennsylvania}
  \country{USA}
}

\author{Qingnan Yang}
\orcid{0000-0002-7125-9867}
\affiliation{%
% \department{Center for Pharmaceutical Policy and Prescribing}
  \institution{University of Pittsburgh}
  \city{Pittsburgh}
  \state{Pennsylvania}
  \country{USA}
}

\author{Kayleigh Adamson}
\orcid{0000-0001-6695-4059}
\affiliation{%
% \department{School of Public Health, Department of Biostatistics}
  \institution{University of Pittsburgh}
  \city{Pittsburgh}
  \state{Pennsylvania}
  \country{USA}
}

\author{Katie Suda}
\orcid{0000-0002-8977-1850}
\affiliation{%
% \department{Center for Pharmaceutical Policy and Prescribing}
  \institution{University of Pittsburgh}
  \city{Pittsburgh}
  \state{Pennsylvania}
  \country{USA}
}

\author{Lu Tang}
\orcid{0000-0001-6143-9314}
\affiliation{%
% \department{Department of Biostatistics}
  \institution{University of Pittsburgh}
  \city{Pittsburgh}
  \state{Pennsylvania}
  \country{USA}
}

\author{Walid Gellad}
\orcid{0000-0002-6992-5197}
\email{walid.gellad@pitt.edu}
\affiliation{%
% \department{Center for Pharmaceutical Policy and Prescribing}
  \institution{University of Pittsburgh}
  \city{Pittsburgh}
  \state{Pennsylvania}
  \country{USA}
}

\author{Adam Perer}
\orcid{0000-0002-8369-3847}
\affiliation{%
  \institution{Carnegie Mellon University}
  \city{Pittsburgh}
  \state{Pennsylvania}
  \country{USA}
}
\email{adamperer@cmu.edu}
%%
%% By default, the full list of authors will be used in the page
%% headers. Often, this list is too long, and will overlap
%% other information printed in the page headers. This command allows
%% the author to define a more concise list
%% of authors' names for this purpose.
\renewcommand{\shortauthors}{Venkatesh Sivaraman et al.}

%%
%% The abstract is a short summary of the work to be presented in the
%% article.
\begin{abstract}
  Drug overdose deaths, including those due to prescription opioids, represent a critical public health issue in the United States and worldwide. Artificial intelligence (AI) approaches have been developed and deployed to help prescribers assess a patient's risk for overdose-related death, but it is unknown whether public health experts can leverage similar predictions to make local resource allocation decisions more effectively. In this work, we evaluated how AI-based overdose risk assessment could be used to inform local public health decisions using a working prototype system. Experts from three health departments, of varying locations and sizes with respect to staff and population served, were receptive to the potential benefits of algorithmic risk prediction and of using AI-augmented visualization to connect across data sources. However, they also expressed concerns about whether the risk prediction model's formulation and underlying data would match the state of the overdose epidemic as it evolved in their specific locations. Our findings extend those of other studies on algorithmic systems in the public sector, and they present opportunities for future human-AI collaborative tools to support decision-making in local, time-varying contexts.
\end{abstract}

%%
%% The code below is generated by the tool at http://dl.acm.org/ccs.cfm.
%% Please copy and paste the code instead of the example below.
%%


%%
%% Keywords. The author(s) should pick words that accurately describe
%% the work being presented. Separate the keywords with commas.

%% A "teaser" image appears between the author and affiliation
%% information and the body of the document, and typically spans the
%% page.
% \begin{teaserfigure}
%   \centering \includegraphics[width=0.8\textwidth]{figures/dashboard_overview.jpeg}
%   \caption{\systemname allows public health experts to compare ML-generated risk scores for geographic regions at the ZIP code or state level. The visualization tool provides (a) a chloropleth map depicting the risk score or other covariates, (b) a slider to show different time periods of data, (c) a time-series plot of variables of interest comparing selected regions against the statewide average, (d) detail panes illustrating the distribution of each variable as well as explanations of the most important features driving the risk score, and (e) descriptions of how the variables have changed over time.}
%   \Description{Todo figure description.}
%   \label{fig:teaser}
% \end{teaserfigure}

\received{January 2024}
\received[revised]{July 2024}
\received[accepted]{October 2024}

%%
%% This command processes the author and affiliation and title
%% information and builds the first part of the formatted document.
\maketitle

\section{Introduction}

Drug overdose is a deadly public health issue. Drug use was involved in over 600,000 deaths globally in 2019, including 125,000 due to opioids alone~\cite{world_health_organization_opioid_2023}.
In the United States, opioids prescribed legally by medical providers are of particular concern, claiming over 17,000 lives in 2021~\cite{CentersforDiseaseControlandPrevention2023}. The opioid crisis has prompted state- and national-level initiatives to create prescription databases that can track individuals at risk of overdose across fragmented medical systems \cite{cdc2021pdmp}.
While these large-scale efforts to monitor and track prescriptions have led to meaningful change in the use of prescription opioids~\cite{pdmp2014mandating,Brown2017}, \textit{local} public health interventions can play a direct role in protecting at-risk individuals by ensuring that life-saving preventative measures are deployed in areas where they are most needed~\cite{albert2011lazarus}.
But local health departments often face steep challenges in resource capacity, data access, and readiness for data-driven technologies~\cite{brownson_building_2018} that have so far hindered their use of centralized prescription data~\cite{cdc2021pdmp}.

% https://www.cdc.gov/drugoverdose/pdf/Leveraging-PDMPs-508.pdf page 45: there is often a disconnect between those who have access to PDMP data and those best positioned to use it to inform public health activities. In most states, intervention activities are largely undertaken by local health departments, while identified PDMP data when available at all are often available only to the state health department

In all 50 U.S. states and the District of Columbia, Prescription Drug Monitoring Programs (PDMPs) have emerged as promising data sources for monitoring overdose risk. These databases, which record every controlled substance prescription made in the state, the prescriber, and who it was prescribed to, have been used to inform state policy and give individual prescribers insight into what other drugs a patient is receiving~\cite{pdmp2014mandating,Brown2017}.
To effectively use PDMP data, prescribers must synthesize many potentially informative variables and their trends over time for a given patient, taking valuable time and effort~\cite{Ferris2019}. Recent literature offers prescribers a clear potential solution to streamline and standardize risk assessment: use artificial intelligence (AI) in combination with PDMP data to predict future risk of overdose~\cite{Ferris2019,Lo-Ciganic2019,Ripperger2021,Gellad2023}. These studies have demonstrated potential utility for \textit{patient}-level decision-making by clinicians, but the potential for use in a \textit{region}-level context remains unexplored.

Despite the impacts that public health agencies can have on the health and safety of communities, few studies in CSCW and HCI have explored how best to support local health departments' work through technology.
Technical efforts to use AI to inform public health decisions bear similarities to the increasing adoption of decision-support systems in other parts of the public sector, including criminal justice~\cite{Angwin2016,Tan2018}, employment services~\cite{ammitzboll_flugge_street-level_2021,Scott2022}, and child welfare~\cite{Vaithianathan2019,Kawakami2022partnerships}. 
However, systems for public health decision-making face distinct challenges to those identified in other public-sector domains.
The use of highly sensitive medical data, for example, limits organizations' abilities to share data and algorithms in a transparent way~\cite{cdc2021pdmp,carroll_visualization_2014}.
Moreover, local health departments have limited capacity due to their numerous responsibilities and, particularly in the rural United States, insufficient funding~\cite{brownson_building_2018,leider_state_2020}.
This excess burden creates opportunities for large-scale data and algorithmic efforts to support under-resourced health departments, but it also places constraints on how and when stakeholders from these organizations can be engaged for design~\cite{backonja_supporting_2022}.
As a result, how local health departments might best leverage emerging algorithmic tools, such as overdose risk assessment models, has thus far been scarcely addressed.

In the present work we aimed to bridge this gap through an exploratory qualitative evaluation embedded within an ongoing project to develop risk prediction algorithms.
Other recent HCI work on designing decision support for experts has either engaged stakeholders \textit{after} an algorithm has been deployed~\cite{Kawakami2022partnerships,Kuo2023} or worked with potential system users from the very earliest stages of design~\cite{HoltenMoller2020,Sendak2020}.
However, in domains such as health where deployment requires extensive validation and expert users are difficult to engage, these approaches are often either too high-risk or infeasible.
Model development efforts instead most commonly focus on scoping out technological capabilities before soliciting user feedback~\cite{gu_lessons_2020,Cai2019,Sivaraman2023}.
By conducting our user evaluation within this relatively technology-centered design process~\cite{yang_re-examining_2020,bly_design_1999}, we could elicit broad perspectives on algorithm-aided decision-making using a functional design probe, enabling more grounded insights.
This approach also suggested ways that such design processes could be improved to better match user needs, particularly when developing AI under the constraints of medical data and the public sector.

Our evaluation was guided by the following question: \textit{What opportunities and pitfalls do local public health officials perceive for AI-based regional risk assessment tools in overdose prevention?} To address it, we built a design probe called \systemname{} to visualize predictions from a model which had been evaluated on historical data but not yet deployed or shown to potential users~\cite{Gellad2023}.
We conducted focus groups with the intended users of such a system: experts and administrators from three local health departments serving populations of varying sizes and demographics in the U.S. state of Pennsylvania. 
Experts were overall receptive to the use of AI-based risk assessment, but they expressed desires for risk visualization to support more than just communicating and explaining model predictions. 
They saw opportunities to use risk prediction tools to communicate and advocate for their priorities across organizations, and to form new collaborations around shared data priorities. 
They also pointed out misalignments between the large-scale, relatively inflexible nature of centralized prescription data and the needs and realities of local public health decision-making. 
These results pose several implications for how to more effectively support local public health officials with AI, particularly when using centralized algorithms to inform local decisions with evolving data needs.

We contribute:
\begin{enumerate}
    \item An \textbf{evaluation of AI-based regional overdose risk prediction} with 11 officials from three local health departments. To ground the evaluation in real data, we designed and built a dashboard to surface predictions and explanations from a working but not-yet-deployed risk assessment model trained on over 5 million patients.
    \item \textbf{Design recommendations} for the future development of AI-driven systems for public health, including explainability, model formulation, and adapting algorithms trained on centralized data to changing local contexts.
\end{enumerate}

\section{Related Work}

\subsection{Data and AI in Public Health}

Given the growing availability of technological systems to track health at scale (a task often termed \textit{public health surveillance})~\cite{Fisher2022,Zeng2020}, public health has seen increased interest in AI. 
HCI literature has contributed both applications and critiques of public health tracking, including novel public health interventions \cite{Balaam2015}, models to predict disease spread based on social media signals \cite{Matero2023,Shahid2020}, and critical discourse around how personal data can or should be used for public health purposes \cite{Thakkar2022,Watson2021}. 
Furthermore, considerable research in data visualization has focused on how visual analytics can support public health and epidemiology applications, as reviewed by \citet{carroll_visualization_2014} and \citet{preim_survey_2020}.
Recently, the COVID-19 pandemic prompted a wave of new tools for tracking and predicting disease spread using geographic visualization \cite{Ghimire2021,Zhang2021,Chande2020}, building on previous work using mapping to track other public health conditions \cite{Maciejewski2011,Allen2016,Miranda2002}. The present work takes inspiration from these geographic risk assessment tools in developing visualizations for overdose risk; however, in this work we focus on the added challenges of communicating an ML-driven risk score.

Despite a large number of technological systems focused on public health, the perspectives of public health decision-makers and administrators themselves have largely been absent thus far from the literature, particularly at the local level. 
For example, many prior data visualization systems for epidemiology demonstrate usability and value through evaluations with domain-expert researchers~\cite{driedger_correction_2007,robinson_combining_2005}, which may yield valuable feedback from a normative perspective yet obscure practical challenges that arise later around adoption and deployment~\cite{carroll_visualization_2014,Fisher2022}.
In a qualitative study by \citeauthor{Morgenstern2021}, public health experts were cautiously optimistic about AI as a tool to help them leverage new data sources without the lag of traditional data collection, but they were also concerned about the risks of selection bias and inequity \cite{Morgenstern2021}. 
Most closely related to our work, \citet{mccurdy_framework_2019} reported a case study on designing a Zika outbreak dashboard for global health workers at a national scale, while \citet{backonja_supporting_2022} developed a dashboard in collaboration with rural health departments to highlight health disparities in their areas.
Our work builds on these findings in a context where an AI system built for a large scale is being considered for deployment into local settings, yielding additional implications for use cases in which priorities at the two levels may not always be aligned.

\subsection{Human-AI Collaboration in the Public Sector}

AI-driven systems have recently been introduced to support decision-makers in a variety of areas in local government, including recidivism prediction \cite{Angwin2016}, child welfare \cite{Vaithianathan2019}, job placement \cite{Scott2022}, and housing allocation \cite{Kuo2023}. These developments have prompted and in turn been influenced by work in CSCW and HCI examining how these systems are used by decision-makers in practice. Many studies consider these systems through the lens of ``street-level algorithms''~\cite{alkhatib_street-level_2019,ammitzboll_flugge_street-level_2021}, which draws a contrast between algorithmic rigidity and the discretion often associated with human-decision makers.
For example, studies with child welfare agencies that use ML-based tools to predict future removal of a child from their home have found that risk predictions measured from administrative data do not align with the notions of risk humans expect them to approximate \cite{Saxena2023}, a finding which workers in these organizations were aware of and compensated for \cite{Kawakami2022partnerships}. Similarly, works by Kuo and Shen et al. \cite{Kuo2023} and \citeauthor{Scott2022} \cite{Scott2022}, on algorithms used in homeless services and in unemployment support, respectively, highlight the tendency of AI-based decision support tools to implicitly view caseworkers' job as accepting some people and denying others (as opposed to advocating for clients and matching people with services). This literature provides a useful starting point for understanding how administrative practices in public health may influence predictive algorithm use at the local level. In another sense, however, algorithms for regional public health are not necessarily best envisioned as ``street-level algorithms'' because they do not purport to replace what is already a highly data-driven decision process~\cite{hoeyer_datafication_2019}. Rather, AI in public health typically aims to provide additional signals to support expert decision-making, creating different challenges for human-AI collaboration that we explore in our study.

Much of the work discussed above has focused on understanding frontline decision-makers' experiences with algorithmic tools \textit{after} they have been deployed and used for some time \cite{Kawakami2022partnerships,Kuo2023,Beede2020}. While this has the advantage of ensuring that decision-makers are familiar with the algorithm and its behavior, it is often difficult to incorporate structural changes suggested by workers into the system's design. A few studies have used early-stage design probes \cite{Kawakami2022} or co-design sessions \cite{HoltenMoller2020} to better understand stakeholders' values alongside algorithm development. Others have developed prototype user interfaces to explore decision-makers' responses to visualizations or algorithmic insights, such as improving the interpretability \cite{Zytek2021} or fairness \cite{Cheng2021} of child maltreatment risk assessments. Our work follows an approach similar to the latter studies by constructing an interactive system based on real data, which we use to understand experts' perceived opportunities for algorithm and interface design \textit{if} such a system were to be deployed.

\subsection{Prototyping and Evaluating Explainable AI Systems}

The design of explainable AI-based systems for decision support has been extensively explored through empirical research in CSCW and HCI literature (reviewed in \cite{lai_towards_2023,rong_towards_2024}).
These studies largely suggest that there is no one-size-fits-all explainability solution for every task.
For example, studies with expert decision-makers have found that feature explanations such as Shapley Additive Explanations (SHAP)~\cite{Lundberg2017} help users understand what data a model is using~\cite{Zytek2021,Cheng2019}, and lead them to perceive the AI as more useful~\cite{Sivaraman2023}.
Yet familiarity with the underlying data and model appears to modulate whether users will gain useful insight from feature explanations~\cite{Wang2020}.
For example, data scientists using feature explanations as a model debugging tool tended to over-rely on misleading explanations in the absence of an in-depth understanding of the data they were working with~\cite{kaur_interpreting_2020}.
For other types of domain experts, feature-based explainability may be less helpful than the simple knowledge that a system has been rigorously validated for their use~\cite{Ghassemi2021,amann_explain_2022}.
Given these often-conflicting findings, there is no clear consensus on how model predictions should be communicated and explained, particularly in under-explored areas such as local public health.
We explored this question through the design and evaluation of our prototype system.

As our evaluation was situated within a model development effort, it forms part of an ongoing, practice-oriented discussion in the literature about how data science teams can most effectively prototype and evaluate possible human-AI interaction designs.
\citeauthor{yang_re-examining_2020} point out several challenges when prototyping and designing AI-driven systems, including helping domain experts understand AI capabilities, rapidly testing human-AI interactions, and combining expertise between AI engineers and end users~\cite{yang_re-examining_2020}. 
Because of these obstacles, previous efforts to design AI-based support systems for experts have often used design concepts without any working system~\cite{yildirim_sketching_2024,Kawakami2022} or evaluations using preliminary but functional systems~\cite{gu_lessons_2020,Cai2019,Sivaraman2023}.
While the former approach is core to human-centered design processes, the latter approach could be described as technology-centered design~\cite{yang_re-examining_2020,bly_design_1999,ries_lean_2011}, in which developers create technological interventions first, then solicit user feedback on them to determine what to build next.
We worked within this technology-centered paradigm primarily because local health officials' workload~\cite{brownson_building_2018,leider_state_2020} made it difficult to engage users in co-design activities without a catalyst for buy-in like a working system. 
However, it also yielded feedback that was more grounded in realistic AI capabilities, and it allowed stakeholders to directly assess the characteristics and limitations of the data.
Since these challenges are likely to exist in other highly-constrained domains, we aimed to identify strategies that future human-AI collaboration projects taking a technology-centered approach can use to more easily adapt to expert feedback when it is available.

\section{Study Context}

Receiving prescription opioids is an important and actionable risk factor for opioid use disorder (addiction) and overdose \cite{CentersforDiseaseControlandPrevention2023}, which constitutes a global public health issue~\cite{world_health_organization_opioid_2023}.
Across the United States, PDMPs aim to mitigate misuse of prescription opioids by capturing information about controlled substance dispensations across jurisdictions. 
They provide a central location for clinicians and other authorized users to review a patient’s current and past prescriptions to assess their risk of opioid use disorder or overdose.
Notably, many state PDMPs were originally designed to help law enforcement investigate drug-related offenses, but their application in public health has become more common as the risks of prescription opioids have become more evident~\cite{cdc2021pdmp,pdmpttac_2018}. 

The nature of PDMPs as statewide prescription tracking systems makes them well-suited for training AI-based systems from a technical perspective.
Namely, they have high data standardization and completeness due to legislative reporting requirements, and they aim to identify individuals consistently across jurisdictions to facilitate tracking~\cite{pdmpttac_2018}.
At the time the algorithm used in our study was developed in Pennsylvania (PA), prescription opioids were a primary driver of overdose-related death~\cite{CentersforDiseaseControlandPrevention2023}, making the PA PDMP an ideal data source to train an algorithm that could be applied in local health departments across the state.
However, more recently an increasing share of opioid-related deaths has been due to synthetic opioids, fentanyl contamination in illegal drugs, and polysubstance combinations of stimulants and opioids that would be less likely to be captured in PDMPS~\cite{adkins_fourth_2024}. 
While prescription drug use continues to be used as an indicator for overdose risk because it is easier to track and has shown good correlation with general overdose \cite{Ferris2019}, the shifting overall trends in overdose serve as a backdrop for the local, time-evolving trends participants described in our evaluation. 

Currently, nearly all states require prescribers to check the PDMP prior to opioid prescriptions, using alerts based on simple measures (high opioid dosage or overlapping prescriptions) or more complicated algorithms \cite{Brown2017,Ripperger2021,Ferris2019,Lo-Ciganic2019}. 
When more complex algorithmic measures of risk have been used in PDMPs, they often lack transparency about their data sources or factors that contribute to elevated risk, and thus provide limited information about opportunities for intervention. 
Therefore, the algorithm used as the basis of our evaluation was among the first of its kind. 
Stakeholders in PA aimed to optimize existing data sources to address the opioid epidemic, as well as to explore the development of a more transparent model that could give prescribers and public health professionals more opportunities for intervention. Therefore, testing how best to visualize and communicate the model predictions for public health was a key objective for our evaluation.
Insights from the evaluation could pave the way for potential future systems that directly integrate into PDMP databases, enabling near-real-time insights on overdose risk. 

\begin{figure}
    \centering
    \includegraphics[width=0.45\textwidth]{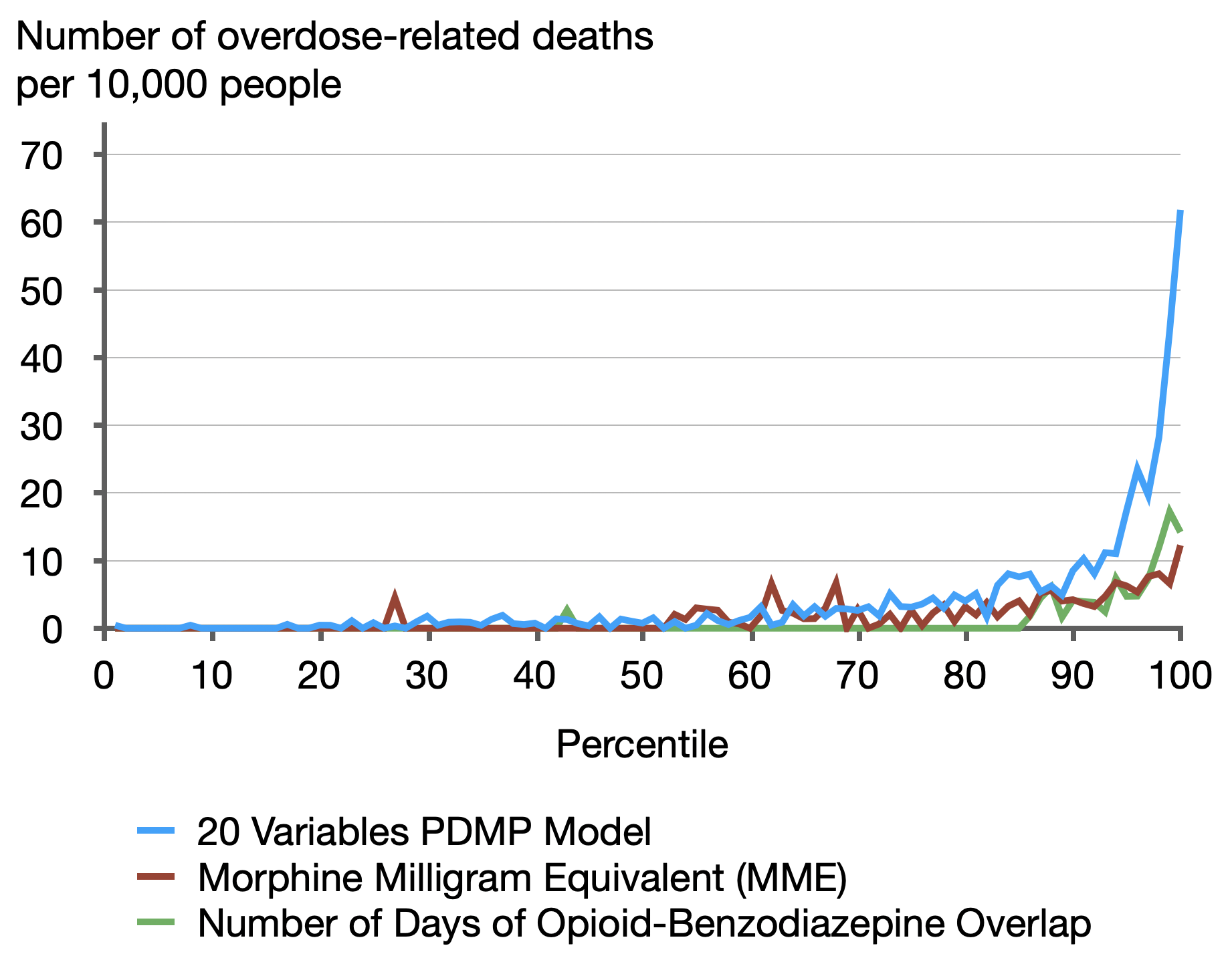}
    \caption{Validation of the PDMP risk prediction model compared to univariate measures of risk currently used by the Pennsylania PDMP. The rate of overdose-related deaths is shown on the $y$-axis by the percentile of each individual's risk, as computed by the model (blue), the patient's morphine milligram equivalent (MME, brown), or number of days of overlapping opioid and benzodiazepine prescriptions (green). Though the rate of fatal overdose is extremely low overall (less than 0.5\%, or 50 deaths per 10,000 people), it shows good correlation with the risk score percentile. Figure adapted from \citep{Gellad2023}.}
    \label{fig:model_validation}
\end{figure}

\section{Design Probe}

We developed a design probe called \systemname{} with the primary aim of helping users directly experience the capabilities and limitations of AI-based risk assessment using real data and model predictions.
To obtain these predictions, we utilized a preliminary overdose risk prediction model developed using PA PDMP data with funding from the Centers for Disease Control and Prevention's (CDC) Overdose Data to Action (OD2A) program.
Below, we provide a brief account of the model's development and validation to contextualize our system design (further details in \cite{Gellad2023}). Then, we describe how we adapted the model's outputs to show regional and time-varying predictions and developed a prototype system with visualizations and AI explanations to enable experts to analyze them.

\subsection{Background: Model Development}

Our evaluation used a risk assessment model, which generates a ``risk score'' for each individual in the cohort. Frequently used in population health and to more efficiently target preventative measures, risk scores are a measure of an individual's risk of experiencing a health outcome within a given time frame--here, fatal opioid overdose within six months. Scores are calculated from a combination of measurable risk factors, are often used to rank or classify members of a population, and can be a way to communicate risk that is more easily understood by laypeople. It is important to note that the risk score at a given time is distinct from the empirical rate of the outcome. Rather, the aggregated risk scores within a region and week represent the risk of \textit{future} overdose in that region, whereas the current overdose rate would be limited to the recent past and would not account for emerging risk factors.

To generate the model, \citet{Gellad2023} partnered with the Commonwealth of Pennsylvania to link PDMP data to unintentional overdose death data and de-identified hospital discharge records, overdose-related emergency department visit data identified via syndromic surveillance, and Emergency Medical Services naloxone administration data. The cohort used to develop the model included all PA residents present in the PA PDMP data from February 2018 to September 2021. This data was linked to unintentional overdose death records for PA and to the State Unintentional Drug Overdose Reporting System (SUDORS) data, which contains detailed records for each fatal overdose. Finally, they linked three other de-identified datasets using the county of residence for each individual, which served to incorporate region-level geographic and socioeconomic risk factors into the model.

We defined the primary outcome of interest as fatal overdose over a 6-month window to make predictions as actionable as possible while ensuring sufficient data availability. After removing individuals who were ineligible for inclusion, the final training/testing cohort included 3,020,748 people with prescriptions and 3,737 deaths within 6 months. The validation cohort followed the same procedures and included 2,237,701 people with prescriptions and 879 deaths within 6 months. Each individual in the dataset was represented using variables computed from PDMP records, including multiple measures of prescription opioid fills including type of opioid, dosage, days’ supply, overlapping medications, and prescription payment information (e.g., cash or insurance pay). The team also created measures describing each individual’s opioid prescribers and pharmacies. In total, 222 features were used for model training and selection; 20 features were selected to train the final model by their feature importance score on the larger models.

After comparing the performance of models developed through a variety of methods, a gradient boosting machine (XGBoost) model was chosen because it performed the best and is computationally efficient. This model had an area under the receiver operating characteristic (AUROC) score of 0.861 on validation data, and a similar AUROC on a cohort of individuals from PA's new PDMP data vendor (who were not included in any previous cohort). Additionally, the model demonstrated a better correlation with mortality rate than indicators of overdose risk currently used by the Pennsylvania PDMP (see Fig. \ref{fig:model_validation}).

\subsection{Aggregating Predictions at Population Scale}

While the model described above captured the risk of fatal overdose for an individual patient at the time of prescription,  for our public health use case we sought to present larger epidemiological trends in overdose risk while protecting individual patient privacy. % public health experts using its outputs to track larger epidemiological trends in overdose risk. For example, they envisioned dividing the state into regions and aggregating the risk score over those regions, thereby helping public health officials prioritize prevention efforts in regions with relatively higher risk levels.
To facilitate this application, we first computed risk scores for 2.8 million patients with records in the PDMP between December 2020 and June 2021. Each risk score output was associated with the ZIP Code Tabulation Area (ZCTA, represented by a 5-digit ZIP code) and county name of the patient's residence. We computed one risk score per patient per week over the 6-month period. Because the risk scores were expressed as estimated probabilities of fatal overdose (which has a very low base rate), we scaled the scores by a factor of 10,000 so that the displayed values would fall within the small integer range, and so each score could be interpreted as the number of expected deaths per 10,000 people. We then aggregated these per-week risk scores at the ZCTA and county levels using two primary metrics: (1) the simple average of the risk scores and (2) the proportion of patients with risk scores in the top decile across the state. The latter metric served to minimize the effects of individual outliers in risk and to fix the baseline statewide score at 10\%, thereby simplifying numerical comparisons.

\begin{figure*}
\centering \includegraphics[width=\textwidth]{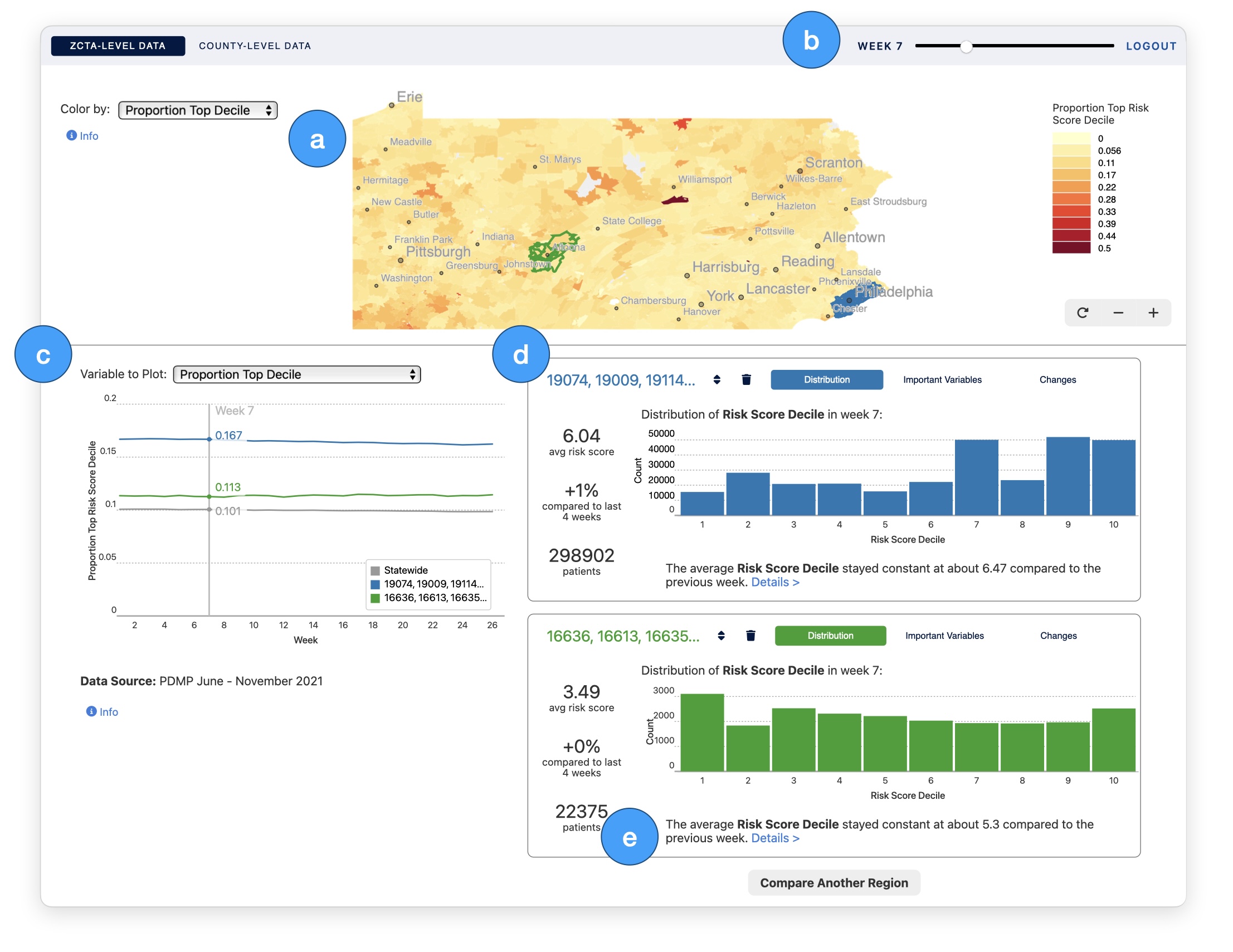}
  \caption{\systemname{} allows public health experts to compare AI-based risk scores for geographic regions at the ZCTA (ZIP code) or county level. The visualization tool provides (a) a choropleth map depicting the risk score or other covariates, (b) a slider to show different time periods of data, (c) a time-series plot of variables of interest comparing selected regions against the statewide average, (d) detail panes illustrating the distribution of each variable as well as explanations of the most important features driving the risk score, and (e) descriptions of how the variables have changed over time. \textit{Note:} the data shown in all screenshots of \systemname{} is outdated and provided for illustration purposes only.}
  \Description{Todo figure description.}
  \label{fig:teaser}    
\end{figure*}
% “Many states are currently using aggregate data for activities like creating data dashboards and compiling reports to provide general information on prescribing trends. Dashboards and other presentations of aggregate data can inform state and local partners to direct resources for targeted education and risk reduction activities, and to provide information to public officials and other interested parties.”

\subsection{Visualization Design}

After generating the spatially- and temporally-aggregated overdose risk predictions described above, we designed a prototype interactive visualization system called \systemname{} to surface these predictions to public health experts. 
Rather than designing novel visualizations that would require local health officials to learn new visual encodings, we drew on empirical evidence and theoretical frameworks of visualization design for consistency with existing tools. We followed a similar process to those described by \citeauthor{gu_lessons_2020} \cite{gu_lessons_2020} and \citeauthor{Sivaraman2023} \cite{Sivaraman2023} in medical AI contexts.
Namely, with the help of governmental stakeholders in Pennsylvania, we identified an initial set of tasks that local public health experts may want to perform using PDMP-based AI risk assessments. Then, we synthesized goals for the specific design of the prototype using existing literature on geographic risk visualization~\cite{Chande2020,Ghimire2021,Maciejewski2011,zakkar_interactive_2017,lechner_towards_2014} and interpretable AI~\cite{Lee2020,gu_lessons_2020,Cai2019}.
Through sketches, mock-ups, and early versions of the working system, we iteratively incorporated feedback from experts in public health and epidemiology (co-authors on this paper) and other stakeholders. 

In its 2021 report on using PDMP data for overdose prevention, the CDC suggests that ``dashboards and other presentations of aggregate data can inform state and local partners to direct resources for targeted education and risk reduction activities''~\cite{cdc2021pdmp}. To facilitate these resource allocation decisions, early conversations with expert stakeholders suggested that our prototype would need to support the following tasks:

\begin{enumerate}[label={\bfseries T\arabic*.}, ref={T\arabic*},itemsep=1ex]
    \item \textbf{Visually highlight regions with high overdose risk.} Stakeholders were particularly interested in using choropleth maps to draw users' attention to high-risk areas, in line with other public health dashboards that use map-based interfaces for navigation~\cite{Chande2020,Ghimire2021,Maciejewski2011}. \label{task:visual-high-risk}
    \item \textbf{Compare regions to each other and to the state.} \citeauthor{few_information_2003} \cite{few_information_2003} calls for interactive dashboards to ``support meaningful comparisons,'' an important goal for \systemname{} since the risk score only carries meaning in comparison to some baseline. For example, officials may want to compare ZCTAs within a county or a county to a neighboring county. In addition to comparing localities, stakeholders emphasized that a statewide average could help local officials easily compare their region to the state overall, providing evidence for whether local intervention was necessary. \label{task:compare-regions}
    \item \textbf{Contextualize AI-generated risk scores with values and trends in raw variables.} Prior work in other public-sector contexts has stressed the importance of algorithmic explainability and transparency to help decision-makers reason about AI predictions \cite{Kawakami2022,Zytek2021}. Similar to these use cases, the PDMP risk prediction model is built on a small number of variables that are often individually meaningful to expert decision-makers, such as the number of opioid prescriptions or percentage of prescriptions filled using cash payments. Experts were very interested in being able to view and compare how these features related to the risk score for one or more regions. \label{task:contextualize-risk-scores}
\end{enumerate}

We then designed specific features of \systemname{} to support these tasks:

\paragraph{Choropleth map to navigate to local analyses} In their synthesis of dashboard design guidelines for public health, \citeauthor{lechner_towards_2014} \cite{lechner_towards_2014} identify the ``customizable, actionable launch pad'' as the most commonly cited design element. Supporting \ref{task:visual-high-risk}, we adopted a map as the ``launch pad'' for \systemname{}. As shown in Fig. \ref{fig:teaser}a, the choropleth map at the top of the page can be colored by a risk variable of interest, including AI-generated risk metrics and other indicators from the model inputs, aggregated at either the ZCTA or county level. The choropleth map colors each ZCTA or country to indicate the average value for all patients in that region.  From this starting point, the user can jump to a time period of interest using the time slider in the top-right corner (Fig. \ref{fig:teaser}b), then select regions by clicking on them on the map. Details about the selected regions are shown in the detail views below the map, including a time-series chart (Fig. \ref{fig:teaser}c) and a detail card showing other explanatory charts about a given selection (Fig. \ref{fig:teaser}d).

\paragraph{Flexible region selection to support comparisons} To construct comparisons (\ref{task:compare-regions}), the user can select one or more regions by clicking on them or performing a lasso selection, which recomputes the aggregations of each metric and updates the detail views accordingly. After creating a selection, the user can click the Compare Another Region button to freeze the existing selection, then make another selection using the same map interactions. The selections are highlighted through different boundary colors on the map. Moreover, each selection is added as a new detail card and a new line on the time-series chart using the corresponding color. Linking selections across all three sections of the dashboard enables public health experts to define and compare regions of interest across both spatial and temporal axes.

\begin{figure*}
    \centering
    \includegraphics[width=\textwidth]{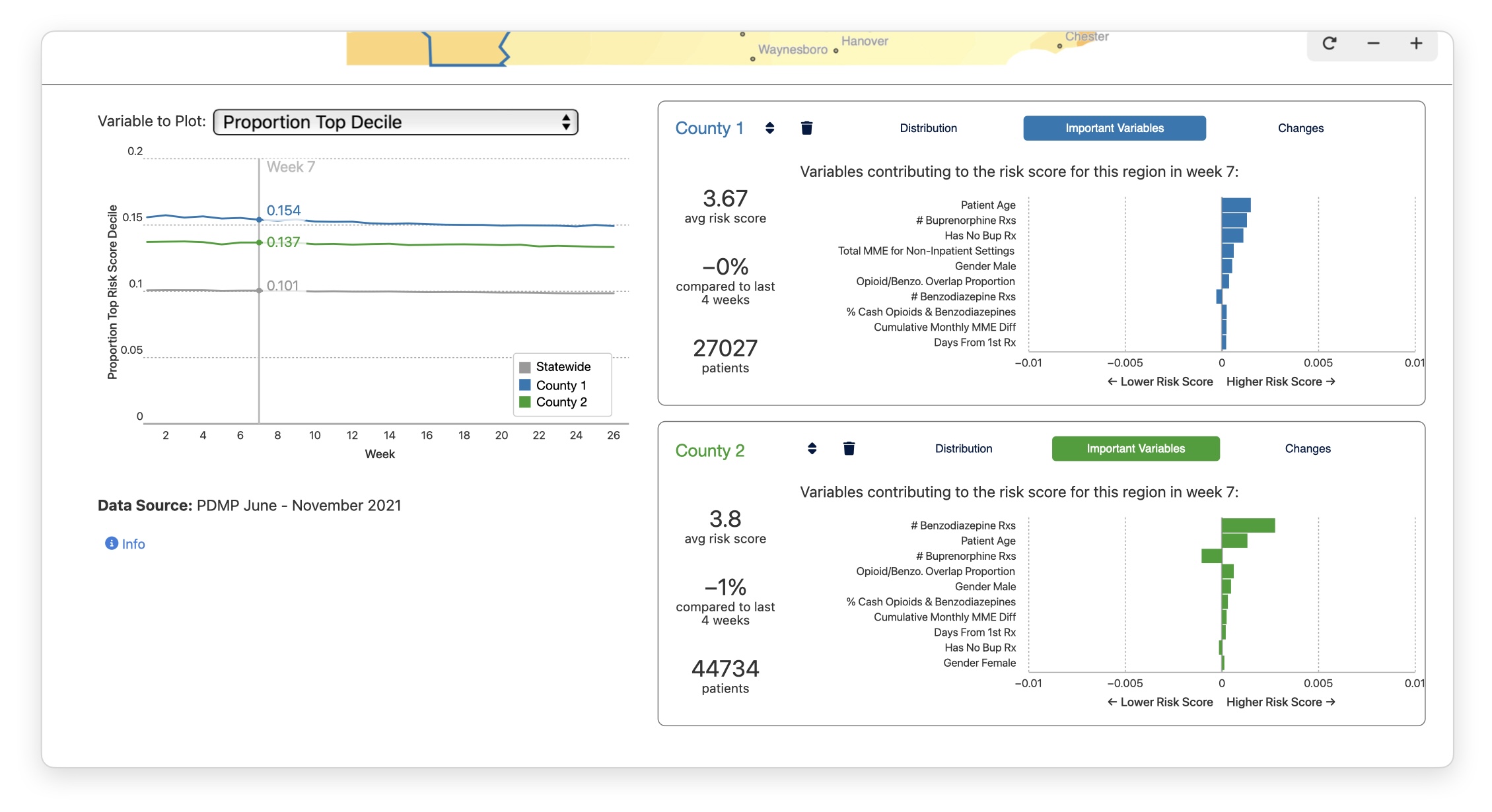}
    \caption{The Important Variables chart shows average feature importance scores for the risk predictions within each selected region and week. Sorting one of the detail cards using the vertical arrow button links the $y$ axis for the charts in the other cards to that region.}
    \label{fig:important_variables}
\end{figure*}

\paragraph{Explanations to link AI outputs to the data features that contribute to them} The Important Variables section in each detail card that the variables that contribute most to predictions in a given region, as shown in Fig. \ref{fig:important_variables}. Compared to other explanation strategies such as counterfactuals~\cite{lee_understanding_2023} or examples~\cite{morrison_impact_2023}, feature importance allowed us to preserve patient privacy and prompt users to look at trends in raw data features, an important part of task \ref{task:contextualize-risk-scores}. We chose to define important variables using SHAP values \cite{Lundberg2017}, since these can be precomputed and aggregated over many instances to produce the average importance for a given region and week. Selecting a bar changes the currently-selected variable throughout the detail views, allowing the user to inspect any trends in the variable's value in the time series plot. Additionally, the user can sort all explanation charts by the SHAP values in one region, enabling an at-a-glance understanding of when variables have very different importance scores in different regions.

\begin{figure}
    \centering
    \includegraphics[width=0.5\textwidth]{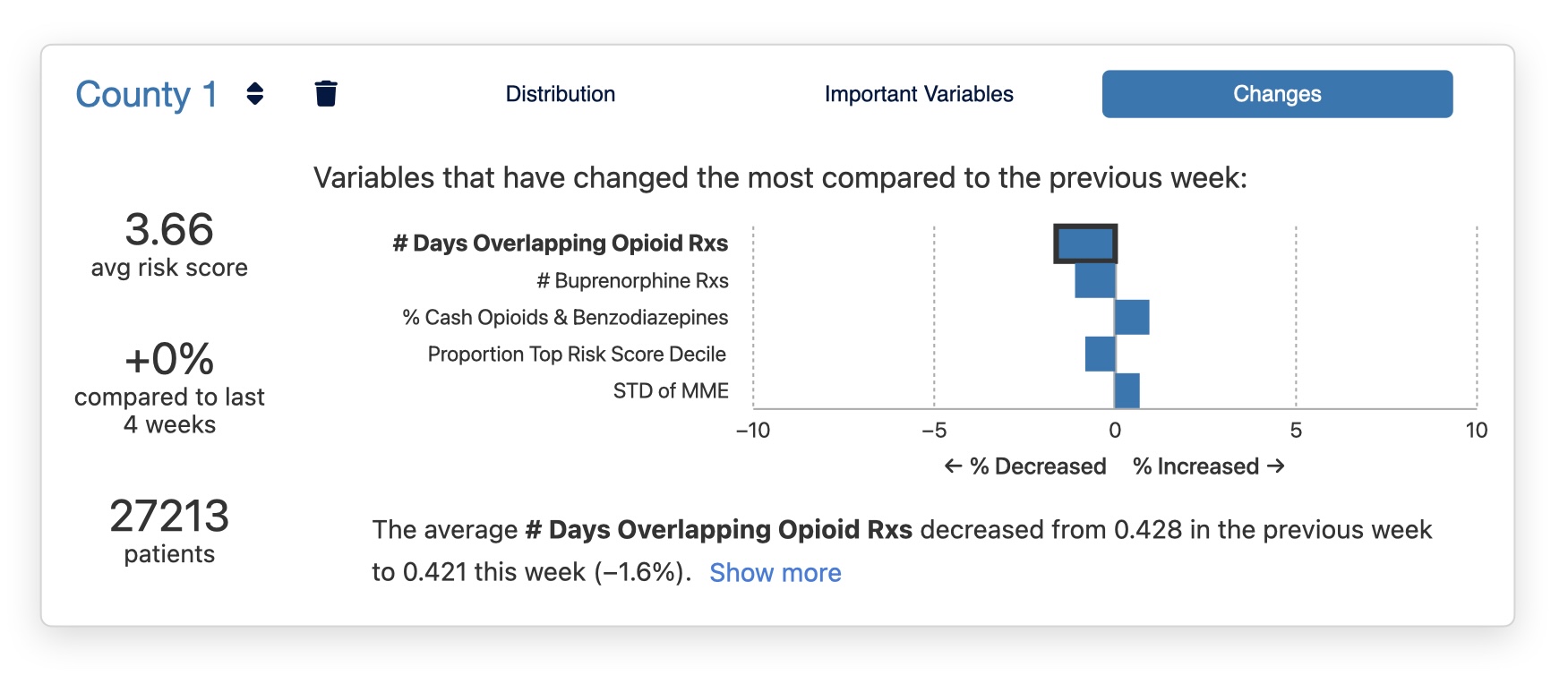}
    \caption{The Weekly Changes pane shows the variables that have changed the most from one week to the next, along with a verbal breakdown of the trend.}
    \label{fig:weekly_changes}
\end{figure}

\paragraph{Direct users to features that have changed meaningfully over time} Discussion with Pennsylvania stakeholders around initial iterations of \systemname{} revealed an interest in not only viewing which factors were important to the risk score for \ref{task:contextualize-risk-scores}, but also which factors were changing meaningfully. Accordingly, we included a Changes pane (Fig. \ref{fig:weekly_changes}) that shows the five variables with the greatest percent change compared to the previous week in a bar chart. To aid the user in interpreting the chart, we included an expandable caption that describes the trend verbally. The text describes whether the selected variable has changed due to incoming patients (patients with new prescriptions), outgoing patients (patients with no new prescriptions in the past 6 months), or changes in the existing patient population.

\paragraph{Preserve patient privacy and mitigate outliers} Development of a dashboard summarizing data from millions of real patients required significant attention to preserving individual privacy. We created an aggregated version of the full dataset that contained no individual identifiers and suppressed all region-week averages and histogram bins that included less than five patients. Additionally, averages were suppressed on the choropleth map for region-weeks containing fewer than 20 patients. This served to reduce the visual saliency of regions that have very few patients but high average risk, while also protecting patients in sparsely-populated areas.

\section{Evaluation Methods}

We used \systemname{} to conduct a series of focus group sessions with officials from three local health departments across Pennsylvania. The following questions guided our evaluation:

\begin{enumerate}
    \item What are local public health officials’ needs for regional risk assessment?
    \item What opportunities do local public health officials see to use AI risk predictions and visualizations in their work?
    \item What are public health officials’ broader attitudes toward a centralized AI-driven tool for risk assessment and visualization?
\end{enumerate}

We used focus groups as the evaluation methodology for both practical and conceptual reasons. Primarily, it was necessary to minimize the time burden on participants as much as possible, since several participants were responsible for large portions of their respective health departments (similar to the recruitment process reported by \citet{Morgenstern2021}). For two of the three counties, visits were scheduled during existing team meetings to ensure that as many team members as possible were able to participate without coordinating additional time. This also afforded the benefit of bringing together members of the ML research team and health department officials with varying types of expertise, which encouraged more wide-ranging conversations. While the focus group setting can sometimes lead to ``groupthink'' effects or the perception of being evaluated, we endeavored to reduce these potential biases by (1) ensuring that other stakeholders (particularly those with funding authority) left the room before the session began, (2) providing participants access to the system beforehand so they could form initial impressions before the meeting, and (3) emphasizing the preliminary nature of the risk assessment tool and welcoming all forms of feedback at the start of the session.

A total of 11 local health department officials from three Pennsylvania counties participated in the evaluation. These health departments varied in size and resource capacity, as well as in the size and demographics of the populations they serve. The participants included public health experts, medical personnel, and people in financial and administrative roles. (Although the latter group was not directly involved in decision-making, they were able to provide useful perspective on the resources and constraints faced by their agencies.) All sessions were conducted on Zoom or Microsoft Teams, and each lasted between 30 and 45 minutes. Participants were given a link to the working prototype and a short description of how it worked, and they were requested to explore the system in advance of the session to familiarize themselves with its capabilities and form initial impressions. During the focus group, the first author led introductions and gave a short demonstration of the system, then followed a semi-structured script consisting of questions about the risk score, how it was presented visually, and how it might interact with participants' roles around overdose prevention. Participants were encouraged to think beyond the capabilities of \systemname{} and envision what types of tools would support their workflow more broadly. Since the sessions could not be recorded, 3-4 team members took detailed notes on participants' responses.

% \begin{table}[]
% \begin{tabular}{rl}
% \toprule
% Participant & Expertise Area                                     \\
% \midrule
% A1          & Epidemiology/Public Health \\
% A2          & Epidemiology/Public Health                         \\
% A3          & Epidemiology/Public Health \\
% A4          & Administrative                                     \\
% A5          & Administrative             \\
% B1          & Medical                    \\
% B2          & Epidemiology/Public Health \\
% B3          & Financial                  \\
% B4          & Administrative            \\
% C1          & Epidemiology/Public Health                         \\
% C2          & Administrative                                    \\
% \bottomrule
% \end{tabular}
% \caption{Summary of participants' roles in their respective health departments (labeled A, B, and C).}
% \label{tab:participant-roles}
% \end{table}

\subsection{Analysis}

We followed an inductive thematic analysis approach to identify themes in participants' reactions and perspectives as captured in our notes. Two of the authors conducted open coding on the notes and resolved any disagreements in the codes and their descriptions through online discussion. Although there were often several versions of each moment captured by different note-takers, the codes for these different accounts were kept separate throughout the process to allow for multiple interpretations. The resulting 157 codes were organized into progressively higher-level themes using affinity diagramming, taking care to group together similar perspectives from multiple sessions when possible. This yielded six over-arching themes, which are presented in the next section.

\section{Evaluation Results}

Broadly, participants across the three focus groups were receptive to the possibilities of integrating risk prediction using a system like \systemname{} into their workflows. At the same time, they envisioned roles for the risk score and dashboard that were outside the scope of the initial design, and they cited critical shortcomings of its underlying data. Additionally, even within just three counties, we heard significant variation in participants' needs for risk prediction, visualization, and decision support. These results, therefore, serve as an initial roadmap for exploring the design space of risk assessment tools for public health.

\subsection{Needs for Regional Risk Assessment}

\subsubsection{Participants currently have little access to predictive modeling or mapping compared to what \systemname{} offers.}\label{theme:high-value-add}

Although the three health departments differed in their size and access to resources, they agreed that the visualization capabilities of \systemname{} were well beyond what they currently use in practice. Participants largely found the tool easy-to-use (A1 [individual 1 from site A], B1, B4), particularly the choropleth as it aligned with previous visualizations they had seen (C1). Participants were particularly excited about the potential to introduce more mapping into their workflow: \textit{``[Our county] is not mapping yet. We do some tracking but a tool like this would be very helpful''} (A). 

Departments A and B, due to their larger capacity and data infrastructure, were accustomed to creating simple visualizations based on locally-relevant data, but not predictive modeling. For example, participants from County B visualized prescribing rates by ZCTA and pharmacy. In contrast, experts from County C (a smaller county) tended to use visualizations much more infrequently, opting for simple charts of fatality rates and naloxone administration when visualizations were required. Respondents from all three counties mentioned that they have previously used quarterly reports and simple dashboards of PDMP data from the state, but that these data sources can be lagged by days to months by the time the reports reach them (A1, A2). As such, A1 noted that \textit{``[\systemname{}] brings more value than anything we have with our PDMP data.''} This theme highlighted that mapping visualization and risk prediction \textit{could} be highly valuable to participants, provided that it fit the expectations and needs they described.

\subsubsection{Participants find it difficult to interpret the variables used in the model and how changes in the variables correlate with trends in risk over time.}\label{theme:interpreting-risk}

Participants from all three health departments found the Important Variables chart useful and interesting, but they also wanted to more rigorously understand the meaning of the twenty variables used in the risk score’s calculation. In one case, for example, A4 observed that the Morphine Milligram Equivalent (MME), a standardized way to measure opioid dosage for a particular patient, would presumably increase risk yet seemed to have a negative feature importance score for the ZCTAs they were analyzing. One explanation for this could be that the variables canonically associated with higher fatal overdose risk are already largely accounted for in decision-making, leading to lower observed overdose rates.

Some participants (B1, B2) used the Important Variables chart in combination with the week slider to examine differences in what factors were contributing to the score over time. However, this led to some confusion among respondents' observations, such as noting that variables might change while the risk score did not: \textit{``If the biggest factor is increasing – which is a good thing – then why isn’t the risk score going down?''} (B1). As B1 pointed out, the model would not be very useful for decision-making without an understanding of how the model’s outputs would change in response to changes in important variables: \textit{``If we are doing what it asks us to but it still isn’t changing, then what do we need to do to make it change?''} 
Statistically, trends in the raw data need not necessarily correspond to changes in the risk score because (a) other variables are not held constant and (b) the predictions shown in \systemname{} are aggregated over many patients.
However, these participants expected the Important Variables to direct them to what was ``driving'' the risk score to increase or decrease (B1, C1). 

Participants from all counties spent time considering the impact of demographics on risk. Notably, age and gender were included in the risk score from the feature selection process, while race and ethnicity were not. Participants were equivocal on whether these demographic variables should inform the risk score and their own decisions, but they agreed that those variables should be available in the dashboard. C1 considered this information more relevant than even other PDMP variables: \textit{``[a trend in] age, gender that is different than what we’re saying, that could be an alert.''} A5 pointed out that although targeting decisions using demographics is ethically fraught, there could be benefits to understanding which groups could benefit from positive interventions: \textit{``Perhaps having the data separated like that would advocate for programs for certain demographics''} such as women or the elderly. In other words, race and ethnicity were considered essential context for understanding risk and making decisions, even if the risk assessment was not directly informed by those variables.

\subsubsection{Participants would find the tool more useful in decision-making if it better reflected geographic characteristics that are relevant to opioid risk.}\label{theme:geographic-characteristics}

All three counties raised questions about the geographic boundaries displayed in the tool, and they expressed interest in being able to manipulate which boundaries were being used for analysis and comparison. ZCTAs, though the most straightforward region definition to use for data collection and aggregation, were perceived as not well-aligned with the areas about which participants make resource allocation decisions (C). For instance, ZCTAs do not necessarily align with county boundaries, so part of a ZCTA may be outside a particular health department's jurisdiction. Some participants suggested instead dividing the map by census tract or neighborhood (A1, C1), which may be more difficult to compute from the source data but would correlate the risk score with more meaningful geographic areas. Similarly, C1 requested that landmarks, such as hospitals, be superimposed on the map, so that they could better contextualize risk in terms of the distance a person might have to travel to get medical help.

While \systemname{} does offer a version of the map aggregated by county, participants rarely used this view because of its lower granularity. (Because all data had to be aggregated at either ZCTA or county level \textit{prior} to inclusion in the dashboard, it was technically infeasible to link the county and ZCTA maps in the current version of the prototype.) Instead of seeing a county-level map, participants suggested that county-level data could be used as a baseline for more fine-grained analysis of ZCTAs (A1, B2). In particular, B2 expressed that this comparison would be more meaningful than the existing statewide baseline: ``I know risk [in county B] is a lot higher compared to the whole state -- if we can compare [ZCTA-level risk] to the whole county instead that would be helpful.'' Helping experts construct context-appropriate baselines is a design opportunity discussed further in Sec. \ref{theme:actionable-score}.

\subsection{Opportunities for Risk Assessment and Visualization}
Having developed an understanding of the risk score and the dashboard, participants reflected on how the tool could inform their decisions as public health professionals and administrators. As described below, these considerations were tightly intertwined with how participants imagined themselves communicating insights from the risk score to others.

\subsubsection{Participants do not see the risk score as actionable unless accompanied by a comparison or threshold that can easily be explained to a stakeholder.}  \label{theme:actionable-score}
Participants across the three counties expressed uncertainty about whether the risk score model could directly inform their decision-making. To C1, the risk score only highlighted regions of high risk that department C was already aware of, so they felt it was better suited for summarizing trends for the semi-regular reviews of historical data conducted by their team. More commonly, participants envisioned the risk score as helping them communicate risk to other stakeholders, such as local government, law enforcement, and medical organizations (A1, B1, B4, C1). But they doubted that the model could be easily presented to non-technical decision-makers: \textit{``What’s the best way to communicate this with a lay audience?''} (A1). C1 wondered, \textit{``How can I give an elevator pitch [for the regional leadership] to take action?... The score would be beneficial, but [is it still preferable] if I need a brief summary to explain to someone how to grasp it?''}

Some participants suggested addressing the difficulty of explaining the risk score to stakeholders by contextualizing it using reference values, comparisons, or correlations with external data. For example, actions could be explicitly defined for risk scores above a certain threshold: \textit{``When you have a score over a certain threshold, it’s easier to tell partners … so [if] for whatever week it's over the average threshold, what does that mean you can do?''} (C1). Alternatively, B1 envisioned presenting risk scores as comparisons against past values for the same region: \textit{``A risk score by itself, if we have a risk score of 12.2 out of 20, imagine how that’s [interpreted]. But if we were 12.2 and now we are 11.8 – so we’re going in the right direction''} (B1). B2 and B4 expressed confusion about what baseline was being used to calculate the risk score, and they suggested that values be compared to nearby counties across different time periods. Regardless of which baseline would ultimately be most appropriate, these perspectives highlighted the ineffectiveness of a single number to estimate “risk” and the importance of comparison to make risk scores meaningful.

\subsubsection{Participants see potential for AI risk predictions to help them collaborate with other organizations to ``connect the dots,'' advocate for resources and disseminate findings.}\label{theme:connect-the-dots}

 In contrast to participants' concern about the usability of the risk score for non-technical stakeholders, they were generally optimistic about the potential for map-based risk visualizations to aid in their communication efforts. For participants from all three health departments, using data to communicate recommendations to stakeholders with a variety of expertise levels was a fundamental need and challenge. For example, C1 explained that \textit{``many of our partners don’t have staff that can handle data, [so we're] always looking for different data that can give them a bigger picture of what’s happening.''} Some participants saw value in being able to use \systemname{} to inform the \textit{``primary prevention and outreach we do... if we suspect there will be a high number of overdoses''} (A1). Similarly, B1 and B4 discussed how detailed regional comparisons could help provide more individualized recommendations to particular parts of their county. The heat map visualizations in the tool were frequently seen as easily exportable and digestible to non-experts: \textit{``They [stakeholders] love to see heat mapping''} (C1).

 However, these participants also perceived broader opportunities for AI-assisted risk assessment to help them move beyond using any single dataset. All three counties, regardless of size and resource capacity, mentioned the difficulties of connecting across different data sources to build an understanding of how their county is doing. This challenge often came up in conversations about data from hospitals, which participants characterized as valuable but often siloed and difficult to access (B, C). For instance, C1 described an effort to conduct outreach with healthcare providers that fell apart due to COVID-19 constraints and misalignments of incentives: \textit{``[We were] providing continuing education practice credits they didn't really need.''} Beyond hospital data, B1 and B4 mentioned that data on housing and rates of acute infectious disease (which could be associated with needle use) were also important and hard to put together. In light of these challenges, participants envisioned systems like \systemname{} as a way to bridge not only presenting the data downstream, but also bringing data \textit{into} the team's purview. By demonstrating the health departments' ability to effectively make use of data, they hoped to improve buy-in from other departments and build relationships that could connect them to other data sources (B, C).
 
\subsection{Perceptions of Standardized Data Analytics and Risk Assessment}

The development of \systemname{} and the model underlying it represents a multi-year effort, and it parallels long-term national trends in the use of PDMP data to guide public health interventions. However, the state of the drug overdose crisis has also evolved over the past several years, leading to potential misalignments between the epidemic that systems like \systemname{} were designed for and the one that is currently being experienced. In all three sessions, participants expressed hesitation around the PDMP and AI-based risk scores derived from it, leading to the following theme:

\subsubsection{Although well-maintained and centralized, the inherent misalignments between PDMP and the real opioid crisis may make the dataset a less useful public health decision aid than direct, local proxy signals.}\label{theme:pdmp-misalignment}

Some participants saw clear value in PDMP data because of its centralization and timeliness (A1, B1, B4). Most datasets that participants worked with regularly were acquired from local collaborators or other governmental health agencies, so they were typically one-off and retroactive (B). Even counties with fairly well-developed data infrastructure were generating new data versions with a few days’ lag. Two of the counties were particularly excited about using a dashboard that could theoretically update even more frequently by leveraging a centralized database. 

However, participants across the three health departments also voiced concerns about using PDMP as the primary dataset underlying the risk score and dashboard: \textit{``I don’t think we can use this to aid all of our public health responses, just because of PDMP limitations''} (B). Specifically, participants described how the priorities of drug overdose prevention work have shifted from prescription drug abuse to mitigating effects of stimulants and more dangerous illegal opioids. As B4 noted, \textit{``people aren’t dying for the same reason as they were 5-10 years ago.''} These participants hesitated to use tools derived from a dataset they saw as systematically under-emphasizing the people they most wanted to offer help to:
\begin{quote}\textit{``With PDMP, we at least know they are seeing a provider. Some people, who are not engaged in care and not getting any prescriptions at all, are at a greater risk }because\textit{ they are not in the PDMP data.''} (B1)
\end{quote}
Meanwhile, to C1, PDMP's focus on drugs obtained from healthcare providers meant that its primary utility was facilitating outreach with hospitals, a task that has been challenging since COVID began. In light of these shifts in the opioid landscape, participants have relied on PDMP data less and less as time progresses – particularly in counties B and C, perhaps because of differences in the primary drivers of overdose in each region.

Given these concerns, participants from these health departments considered how a tool like \systemname{} could diversify the types of data that are used for risk calculation and visualization. For example, participants from county C were interested in overlaying indicators from local offices such as police and emergency services, as well as fatality rates. Connecting across these highly contextual data points was a primary challenge in their work:

\begin{quote}
    \textit{``We're always looking for different data to give us a different picture of what is happening... Is there something new in the drug market? Did a facility close? If there is any way to overlay other data, it would be an added benefit... we can look at this, and we can look at 10 other datasets.''} (C1)
\end{quote}
These types of data pose clear tradeoffs for implementation: they would likely require significant data infrastructure work to collect and centralize, or they may only be applicable to particular regions. On the other hand, if integrated with existing datasets and dashboards they could lead to more directly applicable insights at the local level. B1 emphasized this potential in the case of people who have experienced non-fatal overdoses, which are currently not systematically reported: \textit{“[It would be a game-changer] to be able to galvanize resources post non-fatal overdose. We don’t do enough for these people – what drives change for them?”}

% In 2021, 45 people died each day from a prescription opioid overdose, totaling nearly 17,000 deaths.1 Prescription opioids were involved in nearly 21% of all opioid overdose deaths in 2021.
% https://www.cdc.gov/drugoverdose/deaths/opioid-overdose.html

\section{Discussion}

% opportunities for improved AI to support human-AI collaboration:
% - interactively crafting explanations
% - AI-suggested comparisons
% leveraging local contextual data

As public health departments consider adopting AI-based risk assessment tools to inform their decision-making, it is important we understand how data, models, and interfaces can best complement their current practices. This work explored local public health experts' needs and perceptions for human-AI collaborative risk assessment, revealing several opportunities for future research and system design.
We also reflect on the challenges of designing human-AI predictive systems in the constrained context of public health, and we suggest future research directions that can help teams in such contexts ensure that they are ``getting the right design''~\cite{buxton_sketching} within the bounds of current AI capabilities.

\subsection{Communicating Regional Risk Predictions for Local Public Health}

While decision-maker perspectives on AI systems in the public sector have seen recent interest in the HCI community \cite{Kawakami2022partnerships,Kuo2023,HoltenMoller2020}, the needs of public health professionals have been less extensively discussed. Like other public-sector algorithms, AI tools in public health are often framed as helping experts make more consistent and proactive decisions. Yet these tools often rely on administrative datasets that suffer from systematic biases in which individuals and factors are recorded. Accordingly, decision-makers have expressed skepticism that administrative data of any kind would capture their intended conceptualization of risk, suggesting that algorithmic decision-making may not be well-suited for these tasks. Although participants in our evaluation expressed similar concerns about centralized prescription data, overall they were excited for opportunities to leverage new data sources and predictions to gain a more detailed picture of how their jurisdiction was doing.
More than other public-sector domains described in the literature~\cite{Kuo2023,HoltenMoller2020,Kawakami2022partnerships,ammitzboll_flugge_street-level_2021,Tan2018}, \textbf{AI in local public health needs to account for the fact that a central part of workers' perceived role is making data useful and actionable to partner organizations.} As a result, predictive models can be a useful aid in public health work \textit{if} they help make data more understandable to a non-technical audience.

Towards this goal, participants in our study tried to create narrative rationales of the patterns they observed in \systemname{}'s predictions, and to connect them to model explanations and the local context.
This led to confusion when the feature explanations did not align with the changes in risk score, preventing them from potentially being able to use such narratives when presenting this data to stakeholders.
To overcome this, \textbf{explanations for predictive models in regional, time-varying contexts should be designed for stability}. 
Explanation techniques could ensure that changes in the risk score over time and across areas should be consistent with differences in the explanations, a criterion that has only briefly been discussed in explainable ML literature~\cite{di_martino_explainable_2023}.
Using model explanations as proxies for data trends also creates \textbf{design opportunities for tools that directly help craft data-driven narratives for public health}. 
Visualization systems could mine these explanations for useful patterns, trends, or subsets that could accelerate the process of developing materials to disseminate in the community.
For example, since many stakeholders are accustomed to looking at heat maps to understand differences in risk over space and time, a tool could identify interesting comparisons between regions or weeks and automatically generate smaller maps of these subsets.

Despite the simplicity of a single number that captures the risk associated with all recent prescription data within a region, participants were skeptical that explaining the risk score to stakeholders would be easier than using indicators from the raw data.
One reason for this could be that public health officials would need to vouch for the model's accuracy if using it to justify their decisions; as in clinical decision-making, credible validation results for a model may be an easier path to trust than individually assessing each prediction~\cite{amann_explain_2022,Sivaraman2023}.
If validation studies show that using the risk score is preferable to relying on a few data indicators, \textbf{public health organizations seeking to use these tools may need to associate the risk score with defined policies for action, such as a threshold over which to consider an area ``high risk'' or a target percentage decrease in the risk score to consider an intervention successful}.
These measures may not only improve the actionability of the model, but also lower the communication burden of presenting these decisions to non-technical partner agencies.
On the other hand, rigid policies can also result in mistrust or lack of understanding in AI systems if used to enforce reliance on their outputs~\cite{Kawakami2022partnerships}. \textbf{Model explanations, if designed to present complementary insights from the data, could play a key role in facilitating workers' discretion and providing additional evidence to back up their decisions to stakeholders.}

\subsection{Supporting Dynamic, Local Conceptualizations of Risk}

Our work expands on a common theme in human-centered evaluations of AI-based risk prediction tools, which is that expert decision-makers' notions of risk frequently do not align with those implicitly captured by algorithms.
This can arise because predictive problem formulations are often chosen opportunistically based on what data is most readily available and widely applicable, thereby leading to the highest predictive accuracy~\cite{jacobs_measurement_2021}.
For example, medical billing data~\cite{obermeyer_dissecting_2019} or administrative actions such as a child's removal from their family~\cite{Kawakami2022partnerships} may be available at large scale and conducive to modeling, yet they can create unforeseen misalignments when applied to human-AI decision-making. In the case of fatal overdose, PDMP-based risk factors have been increasingly demonstrated to correlate well with overdose-related death~\cite{Gellad2023,Ripperger2021} even for overdoses not due to prescription drugs~\cite{Ferris2019}.
However, public health crises like the opioid epidemic can evolve rapidly. 
The drivers of opioid overdose are substantially different from the time PDMPs were initially implemented, and from the acquisition of the training data used in our system and others.
Our results suggest that \textbf{even when a prediction model is considered relevant at a large scale, changes in a public health situation over time and across different regions can create perceived misalignments at the local level that require new model formulations to overcome}.

To mitigate these misalignments, future work designing AI systems for public health crises should \textbf{account for the tension between relatively accessible, consistent data sources and the flexibility of local, contextually-specific data.}
The primary promised benefits of AI-based predictive models, namely their accuracy, reproducibility, and generalizability, depend on collecting reliable data from diverse areas in a consistent manner.
In the case of PDMPs and other public health issues, data collection processes typically require top-down regulation mandating that healthcare providers and pharmacies report data such as the quantities of controlled substances dispensed and the identities of their recipients~\cite{cdc2021pdmp}.
In some cases, the existing data is sufficient to capture evolving notions of risk, but prediction models would need to be retrained to account for them. 
For example, increasing use of prescription stimulants can be observed in PDMP data~\cite{padoh_prescribing_2023} though these drugs were not used in our prediction model.
In other cases, however, the most relevant risk signals may not be present in centralized datasets, but rather come from data obtained by local health departments from partner organizations (such as emergency medical services or law enforcement).
As noted in \citeauthor{sun_data_2023}'s and \citeauthor{Thakkar2022}'s accounts of frontline data work, this contextually-rich data may be more heterogeneous in structure, purpose, and incentives behind its collection than statewide databases~\cite{sun_data_2023,Thakkar2022}.
\textbf{It remains an important future direction to investigate how to incorporate these local data sources into risk prediction models,} particularly given the challenges in sharing and standardizing this data.

\citeauthor{Zhang2021}'s study of visualization dashboard creators during the early COVID-19 pandemic \cite{Zhang2021} highlights the challenges of adapting to rapid changes in data and system needs, even in a situation where surveillance data was widely recognized as essential to combat disease spread. In a domain where data infrastructure and centralization is less prioritized, \textbf{the traditional ML lifecycle of data collection, model development, and validation may not be fast enough to appropriately track public health crises.} This challenge could be addressed by lowering the barrier to creation of useful predictive models, such as by providing means to integrate location-specific data into dashboards and risk assessment tools. It may be possible to empower public health experts to create or fine-tune their own models that take advantage of the regional, timely data they have access to, resulting in more actionable risk assessments. However, it may be difficult to validate the success of this approach given the diverse data formats and sources across different health departments. Using AI to instead help users integrate their local data sources into risk visualizations \textit{without} an explicit prediction may be a more intuitive way to aid in combining these disparate signals.

\subsection{Eliciting Rich Expert Feedback During AI Development}

Our evaluation used a predictive model originally designed to support healthcare providers' decisions to prescribe opioids, under the hypothesis that a similar problem formulation could be applied at a regional level. 
Based on our participants' perspectives, however, local health officials might benefit more from predictions for a different cohort of people (e.g., people using illegally-obtained opioids such as fentanyl) using different risk factors and predictive targets (e.g., non-fatal overdoses). 
As \citet{tal_target_2023} and others~\cite{yildirim_how_2022,subramonyam_solving_2022} have discussed, finding a predictive model formulation that is both feasible and relevant is a negotiated process that demands both domain expert input and extensive data science work. 
\textbf{It is therefore critical that model development teams can quickly surface a model’s behaviors for expert user feedback, a task not currently well-supported by data science tools.}
To help teams make best use of experts' time, model building frameworks could aim to make it easier to produce non-technical visualizations of how different model formulations might perform. Then, rather than spending time to present and get feedback on only one predictive task, teams could ask experts to compare across multiple predictive tasks, or even update model formulations on-the-fly.
Particularly in constrained domains such as public health, where data, expertise, and development time are scarce resources, improved techniques for prototyping and evaluating model formulations may be critical to ensuring that AI systems solve the right problems.

\subsection{Limitations} 
The results of our evaluation should be interpreted with the caveat that our scope was limited to public health organizations in three counties within a single U.S. state. 
Furthermore, we were only able to engage these participants after an initial prototype of \systemname{} was built. 
Although our development process was informed by input from other public health experts, incorporating local health departments' perspectives iteratively would likely have allowed us to conduct a deeper exploration of the design space.
Recruiting public health professionals can be challenging given the often overwhelming responsibilities of the job, particularly during the COVID-19 pandemic. 
We hope that future research efforts will continue to enrich our understanding of the data and AI needs of people in these essential roles. 

Another limitation of this work was that due to technical and privacy-related constraints, \systemname{} was constrained to showing the 20 variables used by our risk prediction model \cite{Gellad2023}, which were often combined from basic variables in unintuitive ways for modeling expediency (for example, cumulative monthly difference in prescribed morphine milligram equivalents). 
Future iterations of this system would benefit from showing the underlying primary variables and other PDMP data (such as pharmacy information) that could be relevant to human decision-making at population scale.

\section{Conclusion}

In this work, we developed and evaluated a prototype human-AI collaborative system for public health professionals to assess regional risk of fatal overdose. It is our hope that the HCI community can build on this work by continuing to engage public health experts to learn how AI can best support them through an evolving crisis with changing, diverse data needs. Our results also highlight the importance of flexibility in model formulation throughout the development cycle, including conducting evaluations with working AI prototypes and surfacing data and model limitations to experts early. Exploring and evaluating these challenges earlier in the design process can contribute to the development of more responsible human-AI collaborative risk assessment tools, which can ultimately help guide resources to the people who need them most.

%%
%% The acknowledgments section is defined using the "acks" environment
%% (and NOT an unnumbered section). This ensures the proper
%% identification of the section in the article metadata, and the
%% consistent spelling of the heading.
\begin{acks}
We thank the numerous public health experts at the state and local levels across Pennsylvania for taking time to provide insight and feedback on the dashboard throughout its development. We also thank Terri Washington, Katelyn Morrison, and Donald Bertucci for helping with the feedback collection and analysis process. This work was supported by CDC OD2A (1NU17CE924973), by a National Science Foundation Graduate Research Fellowship (DGE2140739), and by the Carnegie Mellon University Center of Machine Learning and Health.
\end{acks}

%%
%% The next two lines define the bibliography style to be used, and
%% the bibliography file.
\bibliographystyle{ACM-Reference-Format}
\bibliography{references}

%%
%% If your work has an appendix, this is the place to put it.
% \appendix

\end{document}